\documentclass[prb,twocolumn,showpacs,superscriptaddress]{revtex4}
\usepackage{graphicx,psfrag,amssymb,amsmath}
\begin{document}
\title{Spin gap and Luttinger liquid description of the NMR relaxation in carbon nanotubes}
\author{Bal\'azs D\'ora}
\email{dora@pks.mpg.de}
\affiliation{Max-Planck-Institut f\"ur Physik Komplexer Systeme, N\"othnitzer Str. 38, 01187 Dresden, Germany}
\author{Mikl\'os Gul\'acsi}
\affiliation{Max-Planck-Institut f\"ur Physik Komplexer Systeme, N\"othnitzer Str. 38, 01187 Dresden, Germany}
\author{Ferenc Simon}
\affiliation{Institut f\"ur Materialphysik, Universit\"at Wien, Strudlhofgasse 4, A-1090 Wien, Austria}
\affiliation{Budapest University of Technology and Economics, Institute of Physics and Solids in Magnetic Fields 
Research Group, Hungarian Academy of Sciences, P.O. Box 91, H-1521 Budapest, Hungary}
\author{Hans Kuzmany}
\affiliation{Institut f\"ur Materialphysik, Universit\"at Wien, Strudlhofgasse 4, A-1090 Wien, Austria}
\date{\today}

\begin{abstract}
Recent NMR experiments by Singer \textit{et al.} [Singer \textit{et
al.} Phys. Rev. Lett. \textbf{95}, 236403 (2005).] showed a deviation from 
Fermi-liquid behavior in carbon nanotubes with an energy gap evident
at low temperatures. Here, a comprehensive theory for the magnetic
field and temperature dependent NMR $^{13}$C spin-lattice relaxation
is given in the framework of the Tomonaga-Luttinger liquid. The low temperature properties
are governed by a gapped relaxation due to a spin gap %\`a la Haldane
($\sim 30$~K), which crosses over smoothly to
the Luttinger liquid behaviour with increasing temperature.
\end{abstract}


\maketitle

Low dimensional carbonaceous systems, fullerenes, carbon nanotubes
(CNTs), and graphene display a rich variety of exotic states and
strongly correlated phenomena. These include superconductivity in
alkali doped fullerenes \cite{gunnarsson}, quantized transport in single-wall carbon
nanotubes (SWCNTs), 
and massless Dirac quasi-particles showing a halve integer 
quantum Hall-effect in graphene even at room temperature. 
A compelling correlated state of
one-dimensional systems is the Tomonaga-Luttinger liquid (TLL)
state. The TLL state has been suggested to describe the low energy
properties of CNTs with a single shell, the single-wall carbon
nanotubes \cite{eggergog,kane,balents,yoshioka,krotov}.
Transport \cite{bockrath} and photoemission studies
\cite{ishii,rauf} provided evidence for the existence of the TLL
state in SWCNTs. In these studies, power-law behavior of temperature
and bias dependent conductivity and a power-law Fermi edge was
observed, respectively.

Nuclear magnetic resonance (NMR) is a powerful method to
characterize correlated states of materials as it is sensitive to
the density of states near the Fermi edge. For a material with a
Fermi-liquid state,  the temperature
%, $T$ 
dependent spin-lattice relaxation time, $T_1$ follows the so-called Korringa temperature
dependence 
%and 
for which $1/T_1T$ is constant. 
Recently, $^{13}$C enriched SWCNTs were grown inside carbon nanotubes from $^{13}$C enriched fullerenes\cite{simon}.
This allowed a high precision measurement of $T_1$ in small diameter SWCNTs by Singer et al.\cite{singer}.
A tentative fit of the experiments with a 
gapped Fermi liquid type density of states (DOS) indicated overall agreement 
but obvious discrepancies in detail.
When the magnetic field and %$T$ 
temperature dependent data for $T_1$ were fitted with this
phenomenological model a gap at the Fermi surface with $2\Delta
\simeq 40$ K opened already above room temperature, i.e. its $T_c$
is larger than 300 K. This strongly violates the
$2\Delta/\text{k}_{\text{B}}T>3.52$ relation,  thus simple mean field theories are not applicable\cite{gruner}. 
{ Also, the phenomenological description can not account for the strong 
overshoot of $1/T_1T$ when $T$ approaches the gap.}

Here, we analyze the NMR results in the framework of the Luttinger liquid and Luther-Emery liquid pictures 
(interacting one-dimensional electrons without and with a gap, respectively). At high 
temperatures, the former dominates, while the latter accounts for the dominance of a spin gap at low temperatures. 
We show that the temperature and magnetic field dependent
$T_1$ of $^{13}$C can be explained in the TLL scenario with a small spin gap of the order of 30~K.

The most striking difference between Fermi and Luttinger liquids is the anomalous power-law dependence of correlation 
functions at low energies in the latter (including the NMR relaxation rate), with critical exponents changing 
continuously with the 
interaction strength.
This is to be contrasted to a Fermi liquid, where the quasi particle picture (electron) holds, implying 
critical exponents fixed to an integer. In a Luttinger liquid, low energy properties are described by critical phenomena 
of collective modes.

In the following, we construct the low energy theory for CNT with the use of bosonization. We determine the various 
contributions to the NMR relaxation rate at high and low temperatures, and fit the obtained formula to the 
experimental results.

The Hamiltonian describing the kinetic energy of electrons on SWNT is given 
by\cite{eggergog,kane,balents,yoshioka,krotov}
\begin{gather}
H_{kin}=-iv\int
dx\sum_{r \alpha\sigma}r\Psi^+_{r\alpha\sigma}(x)\partial_x \Psi_{r\alpha\sigma}(x),
\end{gather}
where only two transport bands ($\alpha=\pm$) are taken into account, and for each band, there is a right and left 
moving branch ($r=\pm$) with linear dispersion, $\sigma$ stands for the spin.
The forward scattering interaction reads as 
\begin{equation}
H_{FS}=\frac 12 \int dx dy \rho(x) V(x-y)\rho(y),
\end{equation}
where $\rho(x)=\sum_{r\alpha\sigma}\Psi^+_{r\alpha\sigma}(x)\Psi_{r\alpha\sigma}(x)$, $V(x)$ is the Coulomb potential. 
For the moment, we neglect the 
umklapp and backward scattering processes. The Hamiltonian is bosonized via
\begin{equation}
\Psi_{r\alpha\sigma}(x)=\frac{\eta_{r\alpha\sigma}}{\sqrt{2\pi 
a}}\exp\left[i(q_Fr+k_F\alpha)x+i(r\theta_{\alpha\sigma}+\phi_{\alpha\sigma})\right],
\end{equation}
where $k_F$ denotes the undoped Fermi surface of a graphene layer, $q_F$ determines the band filling\cite{eggergog} and $|q_F|\ll k_F$ to ensure the applicability of the low energy continuum approximation. 
$\eta_{r\alpha\sigma}$'s are Majorana fermions, standing for the proper anticommutation relations between different 
branches.
The phase fields are expressed by the symmetric and antisymmetric charge and spin modes ($\delta=\pm$) between the 
valleys of the charge
and spin excitations ($j=c,s$) as 
$2\theta_{\alpha\sigma}=\theta_{c+}+\sigma\theta_{s+}+\alpha\theta_{c-}+\alpha\sigma\theta_{s-}$ and 
$2\phi_{\alpha\sigma}=\phi_{c+}+\sigma\phi_{s+}+\alpha\phi_{c-}+\alpha\sigma\phi_{s-}$.
This leads to the Hamiltonian
\begin{gather}
H_0=H_{kin}+H_{FS}=\nonumber\\
=\sum_{j,\delta} \frac{v_{j\delta}}{2\pi}\int dx 
\left[K_{j\delta}^{-1}(\partial_x\theta_{j\delta})^2+K_{j\delta}(\partial_x\phi_{j\delta})^2\right],
\end{gather}
$v_{j\delta}=v_F/K_{j\delta}$. For long range Coulomb interaction, $K_{c+}\leq 1$ and is 
usually around 0.2, while for the other modes, $K_{j\delta}\geq 1$, but only slightly greater than 
unity\cite{eggergog,giamarchi,rauf,ishii}.

Other interaction terms such as umklapp and backscattering have exhaustively been studied in Refs. 
\onlinecite{eggergog,yoshioka,krotov}, and we refrain from their analyzis here. These parallel closely to the 
investigation of coupled Luttinger liquid chains. Their effect can be summarized as 
follows: at half filling, umklapp scattering generates a finite 
charge (Mott) gap in both 
the symmetric and antisymmetric charge sector, while spin backscattering leads to the opening of a spin gap in the spin 
sectors\cite{balents,kane,yoshioka}. This is
analogous to the spin ladder magnon excitations\cite{kim} found in the strong antiferromagnetic coupling limit.
Thus, the ground state at half filling is a Mott insulator with all  
excitations gapped\cite{yoshiokahalf}.
Away from half filling, the $c+$ sector is gapless as discussed before, while all other sectors remain massive. 
This means, that at low temperatures ($T\ll \Delta$), exponentially activated behaviour is expected\cite{kane,balents}, 
while in the high temperature regime ($T\gg\Delta$), the usual Luttinger liquid behaviour is restored. 
Since the strength of backward and umklapp interactions is inversely proportional to the diameter of the tube,
these become important (and generate a gap) only for narrow tubes\cite{kane,eggergog,giamarchi}, such as the one in Refs. \onlinecite{simon,singer} with an inner tube diameter of 0.7~nm. 

%The above statements apply mainly to long-range Coulomb interactions\cite{eggergog,kane,yoshioka}, which assumes 
%isolated nanotubes. For arrays 
%(ropes), screening becomes important, and Hubbard-like models with short range potentials are 
%invoked\cite{balents,krotov}. Nevertheless, the previous results, including the presence of a spin gap, hold in this 
%case as well.

We start our analysis in the high temperature region to characterize the TLL behaviour first. Hence, we postpone 
the discussion of the gaps for the moment, and concentrate on the 
temperature exponents in the NMR relaxation rate, $1/T_{1}$. 
In general, $1/T_1$ measures the local dynamics of the spins, and it is related to the transverse spin susceptibility, 
which in the case of nanotubes, reads as
\begin{equation}
\chi_{\perp}(i\omega_n)=-\int\limits_0^\beta\textmd{d}\tau e^{i\omega_n\tau}\langle
T_\tau S^+(x=0,\tau)S^-(x=0,0)\rangle,
\end{equation}
where $S^\pm(x)=\sum_{r\alpha r^\prime\alpha^\prime}\Psi^+_{r\alpha \pm}(x)\Psi_{r^\prime\alpha^\prime \mp}(x)$.
From this, after analytic 
continuation to real frequencies, the NMR relaxation time is
\begin{equation}
(T_1T)^{-1}\sim \lim_{\omega\rightarrow 0}\frac{\textmd{Im}\chi_{\perp}(\omega)}{\omega}.
\end{equation}
In general, there are four different length scales\cite{eggergog} in $S^\pm(x)$, and the corresponding wavevectors are
1. $q_1=0$, 2. $q_2=2k_F$,  3. $q_3=2q_F$ and 4. $q_4=2(k_F\pm q_F)$. These show up in the term to $S^\pm(x)$ at 
1. $\alpha=\alpha^\prime$, $r=r^\prime$, 
2. $\alpha=-\alpha^\prime$, $r=r^\prime$, 
3. $\alpha=\alpha^\prime$, $r=-r^\prime$ and 
4. $\alpha=-\alpha^\prime$, $r=-r^\prime$, respectively.

\begin{table}[h!] 
\centering

%\begin{tabular}{|c|c|c|c|}
%\hline
%\hline
%$q=0$ & $q=k_F$ & $q=q_F$ & $q=k_F\pm q_F$ \\ \hline 
%$\gamma_1$ & $\gamma_2$  & $\gamma_3$ & $\gamma_4$ \\ \hline
%$\dfrac{1}{K_{s+}}+K_{s+}+\dfrac{1}{K_{s-}}+K_{s-}$ & $\dfrac{1}{K_{c-}}+K_{c-}+\dfrac{1}{K_{s+}}+K_{s+}$ & 
%$K_{c+}+K_{c-}+\dfrac{1}{K_{s+}}+\dfrac{1}{K_{s-}}$ & 
%$K_{c+}+\dfrac{1}{K_{c-}}+\dfrac{1}{K_{s+}}+K_{s-}$ \\
%\hline
%\end{tabular}

\begin{tabular}{|c|c|c|c|}
\hline
\hline
$i$ & $q_i$ & $\alpha$, $\alpha^\prime$, $r$, $r^\prime$ & \textmd{exponent} ($\gamma_i$) \\ 
\hline
1& $0$& $\alpha=\alpha^\prime$, $r=r^\prime$ &   $\dfrac{1}{K_{s+}}+K_{s+}+\dfrac{1}{K_{s-}}+K_{s-}$ \\ 
\hline
2& $2k_F$ & $\alpha=-\alpha^\prime$, $r=r^\prime$  & $\dfrac{1}{K_{c-}}+K_{c-}+\dfrac{1}{K_{s+}}+K_{s+}$ \\ 
\hline
3& $2q_F$ & $\alpha=\alpha^\prime$, $r=-r^\prime$  & $K_{c+}+K_{c-}+\dfrac{1}{K_{s+}}+\dfrac{1}{K_{s-}}$ \\ 
\hline
4& $2(k_F\pm q_F)$ & $\alpha=-\alpha^\prime$, $r=-r^\prime$ & 
$K_{c+}+\dfrac{1}{K_{c-}}+\dfrac{1}{K_{s+}}+K_{s-}$ \\
\hline
\end{tabular}

\caption{The long time exponents of the different contributions to the transverse spin susceptibility ($\sim 
t^{-\gamma_i/2}$). These also determine the temperature dependence
of the NMR relaxation rate as $(T_1T)^{-1}\sim T^{\gamma_i/2-2}$.
\label{tab1}}%
\end{table}%

In all cases, the corresponding contribution to the susceptibility decays at long 
times as $\sim 1/t^{\gamma_i/2}$ with the exponents listed in Table \ref{tab1}.
Based on Ref. \onlinecite{eggergog}, these can be approximated by
$\gamma_1\approx \gamma_2\approx 2+1/K_s+K_s$ and $\gamma_3\approx\gamma_4 \approx K_c+2+1/K_s$ with $K_s=K_{s+}$ and 
$K_c=K_{c+}$. In the most general case with broken spin rotational symmetry, all these exponents usually differ from 
each 
other. However, from realistic models\cite{kane,balents,yoshioka}, one can conjecture their values as $K_{s-}=K_{c-}=1$.
In the presence of external magnetic field, which couples to the $s+$ field and breaks the spin rotational invariance, 
marginal 
interactions change the exponents\cite{giamarchi}, hence $K$'s can differ 
from their zero field value. This follows from the exact Bethe-Ansatz solution of the Hubbard model in magnetic 
field\cite{frahm} as well.
Based on all this, a simple power counting determines the temperature exponents in the NMR relaxation rate as
$\sim T^{\gamma_i/2-2}$.

Having determined the high temperature Luttinger liquid behaviour, we now turn to the evaluation of the response 
function at low temperatures where the gaps are dominating. In 
general, at $T=0$, all the correlation functions 
containing a gapped field acquire a factor $\sim \exp(-\Delta\sqrt{\tau^2+v^2x^2})$, which suppresses exponentially the 
long time-long distance power-law Luttinger liquid behaviour\cite{voitgap,wiegmann}. This picture is further 
corroborated by the existing exact 
solutions of the attractive Hubbard model along the Luther-Emery line ($K=1/2$), where strongly interacting gapped 
bosons are mapped onto non-interacting 
fermions\cite{giamarchi,nersesyan}, i.e. the non-interacting massive Thirring model\cite{japaridze}. In some cases, 
this mapping also 
allows for an explicit evaluation of certain correlation functions, such as the finite momentum transverse spin 
susceptibility\cite{japaridze,voitgap}.

At finite temperatures, this simple picture needs to be modified. From the exact solution of gapped systems such as the 
sine-Gordon model\cite{giamarchi,nersesyan}, it is well established that different type of excitations contribute to higher frequencies and temperatures\cite{levitov}. The 
simplest ones are the solitons and antisolitons (kinks), which, when attract each other, form breathers.
Their contribution can systematically be analyzed based on the "Form Factor Bootstrap Approach"\cite{essler,orgad}, 
which uses the integrability of the underlying model (in the case of the Hubbard model, for example) to determine the matrix elements of the appropriate operators between the 
ground state and excited states. In this case, a simple exponential term cannot account for their variety.

The explicit determination of these matrix elements is certainly beyond the scope of the present investigation. Instead, 
we utilize our knowledge about these excitations. At temperatures comparable to $\Delta$, they start to 
contribute 
significantly to the response functions. This situation can be mimicked by retaining our simple exponential factor in the 
correlation functions, and adjusting the gap at each temperature to give reasonable contribution. In other words, this 
means a $\Delta(T)$ gap function.
In spirit, this approach is identical to the self-consistent harmonic approximation\cite{mocanu}, where one 
replaces the 
complicated 
interaction term (such as the cosine in the sine-Gordon model) by simpler quadratic ones, and optimizes its coefficient 
to minimize the free energy of the system\cite{maki}. Such an approach for the sine-Gordon model leads to a mass which 
monotonically decreases with temperature, while above certain $T\sim\Delta(T=0)$, the potential generating a gap for a 
field is completely washed due to the fluctuations of the field itself\cite{maki}.
Above these temperatures, the mass practically disappears. Such an approach gives very accurate results for the 
dependence of the gap 
on model parameters at zero temperature\cite{giamarchi,mocanu}.

At finite temperatures, contributions to the real time transverse spin correlator read as
\begin{equation}
\chi_{\perp}(t)\sim -\dfrac{\left(T 2a\pi/v\right)^{\gamma_i/2}}
{\left|\sinh(\pi Tt)\right|^{\gamma_i/2}}\exp\left(-i\frac{\gamma_i\pi}{4}-i\Delta(T)|t|\right),
\end{equation}
which reduces to the standard Luttinger liquid correlator in the absence of the gap. 
Here $\Delta$ is the sum of the existing gaps. At half filling, all sectors are gapped\cite{yoshiokahalf}, hence 
$\Delta$ is their sum. Away from half filling, which is presumably the case in 
Ref. \onlinecite{singer}, the spin sectors are fully gapped\cite{yoshioka}, and the $\Delta$ is dominated by their 
contribution. 
From this, by Fourier transformation, we can evaluate the frequency dependent retarded correlation functions, from which 
we can determine the NMR relaxation rate as
\begin{equation}
(T_1T)^{-1}=A\left(\frac{2a\pi}{v}\right)^{K}T^{K-2}\textmd{Im}F(T),
\label{nmrt1}
\end{equation}
where
\begin{gather}
F(T)=\exp\left(\frac{iK\pi}{2}\right)\dfrac{\Gamma(1-K)\Gamma\left(-\dfrac{i\Delta(T)}{2\pi
T}+\dfrac
K2\right)}
{\Gamma\left(-\dfrac{i\Delta(T)}{2\pi T}-\dfrac K2+1\right)}\times\nonumber\\
\times\left[\Psi\left(-\frac{i\Delta(T)}{2\pi T}-\dfrac K2+1\right)-\Psi\left(-\frac{i\Delta(T)}{2\pi T}+\frac 
K2\right)\right],
\end{gather}
and $K=\gamma_i/2$, $A$ is a proportionality factor, which is determined by the contribution of the respective fermionic 
fields and cannot be obtained by the method used here, $\Gamma(x)$ and $\Psi(x)$ are the gamma and digamma functions, 
respectively. At high temperatures, Eq. \eqref{nmrt1}  exhibits a $T^{K-2}$ 
behaviour\cite{yoshiokanmr}, as is 
expected from Luttinger 
liquids\cite{nersesyan,giamarchi}. Based on the possible values of Luttinger liquid parameters, the contributions with 
$\gamma_3$ and $\gamma_4$ are the most divergent ones at low temperatures due to the presence of the $c+$ mode, hence we 
will concentrate on them in the followings. They lead to $K=1+(K_c+1/K_s)/2$.

\begin{figure}[h!]
%\psfrag{x}[t][b][1][0]{$T$~(K)}
%\psfrag{y}[b][t][1][0]{$(T_1T)^{-1}$~(sK)$^{-1}$}
{\includegraphics[width=7cm,height=7cm]{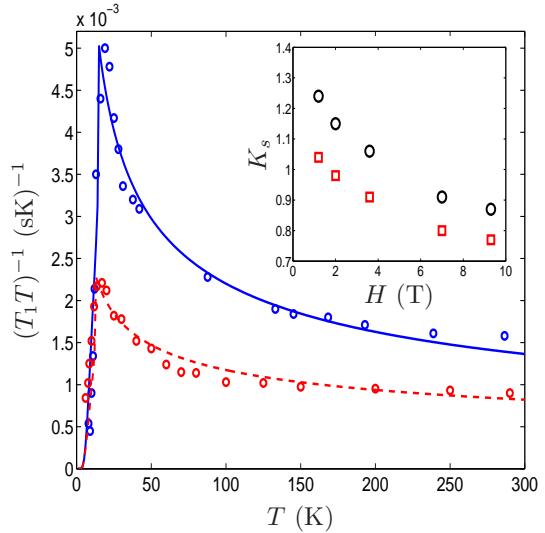}}

%\vspace*{-65mm}\hspace*{25mm}
%\psfrag{x}[t][b][1][0]{$H$~(T)}
%\psfrag{y}[b][t][1][0]{$K_s$}
%{\includegraphics[width=34mm,height=34mm]{ksvsh.eps}}

%\vspace*{31mm}
\caption{(Color online) The fit of Eq. \eqref{nmrt1} to the experimental data (open circles) of
Ref. \onlinecite{singer} at $H=3.6$~T (red dashed line) and 9.3~T (blue line). The inset shows the deduced $K_s$
exponents as a function of the field for $K_c=0.2$ (black circles) and 0.05 (red squares).
\label{fit1}}
\end{figure}

In Fig. \ref{fit1}, we fit our Eq. \eqref{nmrt1} to the experimental data on the inner wall of metallic double-wall 
carbon nanotubes as a function of temperature at magnetic fields $H=3.6$~T and 9.3~T. 
To increase the strength of the NMR signal, selective enrichment of the inner shells using $^{13}$C isotope was 
performed, hence mainly the properties of the inner tube  has been probed by the magnetic resonance measurements. 
Here, we take
$\Delta=\Delta_0(1-T/T_c)^{1/4}$, which is very close to that found in the self consistent solution of the soliton energy in the sine-Gordon model\cite{maki}. With this, we can nicely account for the various complicated excitations of our model. 
Obviously, we expect a smooth crossover of $\Delta$ as the temperature increases through $T_c$ to a 
small value ($\Delta(T)/T\ll 1$) rather than a sharp drop to zero, but for computational purposes, 
we use the above sharp form. 
Our results are robust with respect to variations of the gap function, e.g. the quality of the fitting remains the same 
by changing the exponent, 1/4, within the 
range 0.1 - 0.7,  since only a finite gap a
low temperatures ($T\ll T_c$) and a tiny gap at high temperatures is required ($T\gg T_c$).
For the fits, we use $T_c\approx 13$~K and $\Delta_0\approx 32$~K. 
Theoretically\cite{maki}, the critical temperature, at which the cosine potential practically disappears from the 
sine-Gordon model, is roughly determined as $T_c\approx \Delta(0)/e$, in perfect agreement with our fitting.
Above $T_c$, the 
gap does not make itself felt any more. For the Luttinger liquid parameters, we assume $K_c=0.2$, in accordance with 
Refs. \onlinecite{eggergog} and \onlinecite{yoshioka}, and deduce $K_s=1.07$ for $H=3.6$~T and $K_s=0.87$ for 
$H=9.3$~T, which vary with the 
field but stay close to unity as expected\cite{giamarchi}. 
For the sake of completeness, we give the numerical prefactors, which are found to be $A=1.4\times 10^4$~K/s, 
while upon reinserting original units, $k_B 2a\pi/\hbar v=10^{-4}$~1/K, and $v$ can be taken roughly as $v_F$. 
The latter falls into the same order of magnitude, if we identify $a$, as a rough estimate, to be of the order 
of the honeycomb lattice constant ($\sim  2.46$~\AA), and $v_F\sim 10^6$~m/s, which again certifies our approach.
Since the NMR relaxation rate has been measured at 290~K for different fields\cite{singer}, we can also determine the 
field 
dependence of the $K_s$ exponent, which is shown in the inset of Fig. \ref{fit1} for two different, field independent 
values of $K_c$. It stays close to 1 and decreases with field. The other exponents $K_{c\pm}$ and $K_{s-}$ can also 
vary with the field, but since no exact solutions are available for the Luttinger liquid exponents of carbon nanotubes 
unlike for the 
Hubbard model\cite{frahm}, we choose the simplest possible approach of assigning all the field dependence to the 
${s+}$ field. It is evident from the figure, that the larger $K_c$, the larger $K_s$, hence a suitable increase of the 
former in magnetic field can result in an increment of the latter.
Given the simplicity of our scheme, the obtained fits are excellent and the parameters are 
reasonable.
Finally, we mention the possibility, that upon doping, the properties of the $c+$ mode are expected to change 
(e.g. $K_{c+}$) similarly to the change of the $s+$ mode in a magnetic field.

In conclusion, we have analyzed a recent experimental data\cite{singer} of the NMR relaxation rate of CNT. With the 
use of bosonization, we find that the low temperature region is well described by a spin gapped Luther-Emery liquid.
Relaxation is dominated by exponentially activated behaviour with a gap $\sim 30$~K.
This crosses over to the usual Luttinger liquid picture at high temperatures compared to the gap, with characteristic 
non-integer power-law exponents. 
Finally, we predict, that the observed spin gap can significantly be reduced and even completely suppressed through a 
quantum phase 
transition by experimentally accessible magnetic fields ($\sim 25-30$~T) due to its small value, 
similarly\cite{giamarchi,chaboussant} to 
Cu$_2$(C$_{5}$H$_{12}$N$_{2}$)$_2$Cl$_4$.

%\begin{acknowledgments}
The authors acknowledge useful discussions with A. Virosztek.
Supported by the FWF project I83-N20 (ESF IMPRESS) and by the
Hungarian State Grants (OTKA) No. TS049881, F61733 and NK60984. FS
acknowledges the Zolt\'{a}n Magyary and the Bolyai programmes for
support.
%\end{acknowledgments}

\bibliographystyle{apsrev}
\bibliography{luttref}
\end{document}